\documentclass[english,prl,aps, superscriptaddress,twocolumn,longbibliography]{revtex4-1}
\usepackage[T1]{fontenc}
\usepackage[latin9]{inputenc}
\usepackage{mathrsfs}
\usepackage{amstext}
\usepackage{amsmath}
\usepackage{graphicx}
\usepackage{esint}

\makeatletter
\usepackage[colorlinks,citecolor=blue,linkcolor=blue]{hyperref}

\makeatother

\usepackage{babel}
\begin{document}
\title{Experimental validation of the $1/\tau$-scaling entropy generation
in finite-time thermodynamics with dry air}
\author{Yu-Han Ma}
\address{Beijing Computational Science Research Center, Beijing 100193, China}
\address{Graduate School of China Academy of Engineering Physics, No. 10 Xibeiwang
East Road, Haidian District, Beijing, 100193, China}
\author{Ruo-Xun Zhai}
\address{Beijing Normal University, Beijing 100875, China}
\address{Graduate School of China Academy of Engineering Physics, No. 10 Xibeiwang
East Road, Haidian District, Beijing, 100193, China}
\author{Chang-Pu Sun}
\address{Beijing Computational Science Research Center, Beijing 100193, China}
\address{Graduate School of China Academy of Engineering Physics, No. 10 Xibeiwang
East Road, Haidian District, Beijing, 100193, China}
\author{Hui Dong}
\email{hdong@gscaep.ac.cn}

\address{Graduate School of China Academy of Engineering Physics, No. 10 Xibeiwang
East Road, Haidian District, Beijing, 100193, China}
\begin{abstract}
The second law of thermodynamics can be described as the non-decreasing
of the entropy in the irreversible thermodynamic process. Such phenomenon
can be quantitatively evaluated with the irreversible entropy generation
(IEG), which was recently found to follow a $1/\tau$ scaling for
the system under a long contact time $\tau$ with the thermal bath.
This scaling, predicted in many finite-time thermodynamic models,
is of great potential in the optimization of heat engines, yet remains
lack of direct experimental validation. In this letter, we design
an experimental apparatus to test such scaling by compressing dry
air in a temperature-controlled water bath. More importantly, we quantitatively
verify the optimized control protocol to reduce the IEG. Such optimization
shall bring new insight to the practical design of heat engine cycles.
\end{abstract}
\maketitle
\textit{Introduction}\textit{\emph{-}} Heat engines, converting heat
into useful work, have important practical applications and attract
a wide range of research interests in both classical and quantum thermodynamics
\citep{Carnoteff,Esposito_2009,Campisi_2011,Pekola_2015,Vinjanampathy_2016,Binder2018}.
In classical thermodynamics, the Carnot theorem \citep{Carnoteff}
limits the maximum efficiency of heat engines with the well-know Carnot
efficiency $\eta_{\mathrm{C}}=1-T_{\mathrm{c}}/T_{\mathrm{h}}$ ,
where $T_{\mathrm{c}}(T_{\mathrm{h}})$ is temperature for the cold
(hot) bath. Unfortunately, achieving such efficiency is typically
accompanied by a vanishing output power due to the infinite long operation
time in a quasi-static thermodynamic process \citep{Carnoteff,Bender2000,Humphrey2002,QTquanQH,YHMaQPTHE}.
The futility of such heat engine with vanishing power has pushed to
design finite-time cycle to achieve high efficiency while maintaining
the output power \citep{CA,MinEntropyProd,SekimotoJPSJ,BroeckPRL2005,Tu2008JPhysAMathTheor41_312003,EspositoPRL2010}.
For such design, the quantitative evaluation of the irreversibility
is the key for optimization \citep{CA,andresen1984thermodynamics,Chen1994The,BroeckPRL2005,Tu2008JPhysAMathTheor41_312003,EspositoPRL2010,linearModelHulobec,Holubec2017,Binder2018}.
The trade-off relation between power and efficiency \citep{tradeoffholubec,TradeoffrelationShiraishi,CavinaPRLtradeoffrelation,Constraintrelationyhma}
are significantly determined by the relation of irreversible entropy
generation (IEG) on the control time $\tau$. In the near-equilibrium
region, the IEG in a finite-time isothermal process is found inversely
proportional to the process time $\tau$, namely the $\mathscr{C}/\tau$
scaling with the coefficient $\mathscr{C}$. Such scaling has been
discovered in different finite-time thermodynamic models \citep{ZCTuCPB},
such as endo-reversible model \citep{CA,endoreversibleHE,MinEntropyProd,Sahin1996},
linear model \citep{BroeckPRL2005,Wang2012,Stark2014,linearModelHulobec},
stochastic model \citep{EffBrownianHE,SeifertEPLStochatic,IzumidaEPLStochatic,StochasticThermodynamics,Dechant2015},
and low-dissipation model \citep{EspositoPRL2010,Tomas2012,Constraintrelationyhma}.
Moreover, the scaling relation has been established not only for the
classical working substance \citep{MinEntropyProd}, but also for
quantum working substance \citep{CavinaPRLtradeoffrelation,Constraintrelationyhma,yhmaoptimalcontrol}.
The coefficient $\mathscr{C}$ is determined by the statistical properties
of the working substance and the heat bath, and has recently been
proved to be related to the way that the working substance being manipulated
\citep{Gong2016PRL,yhmaoptimalcontrol}.

\begin{figure}
\begin{centering}
\includegraphics[width=8.5cm]{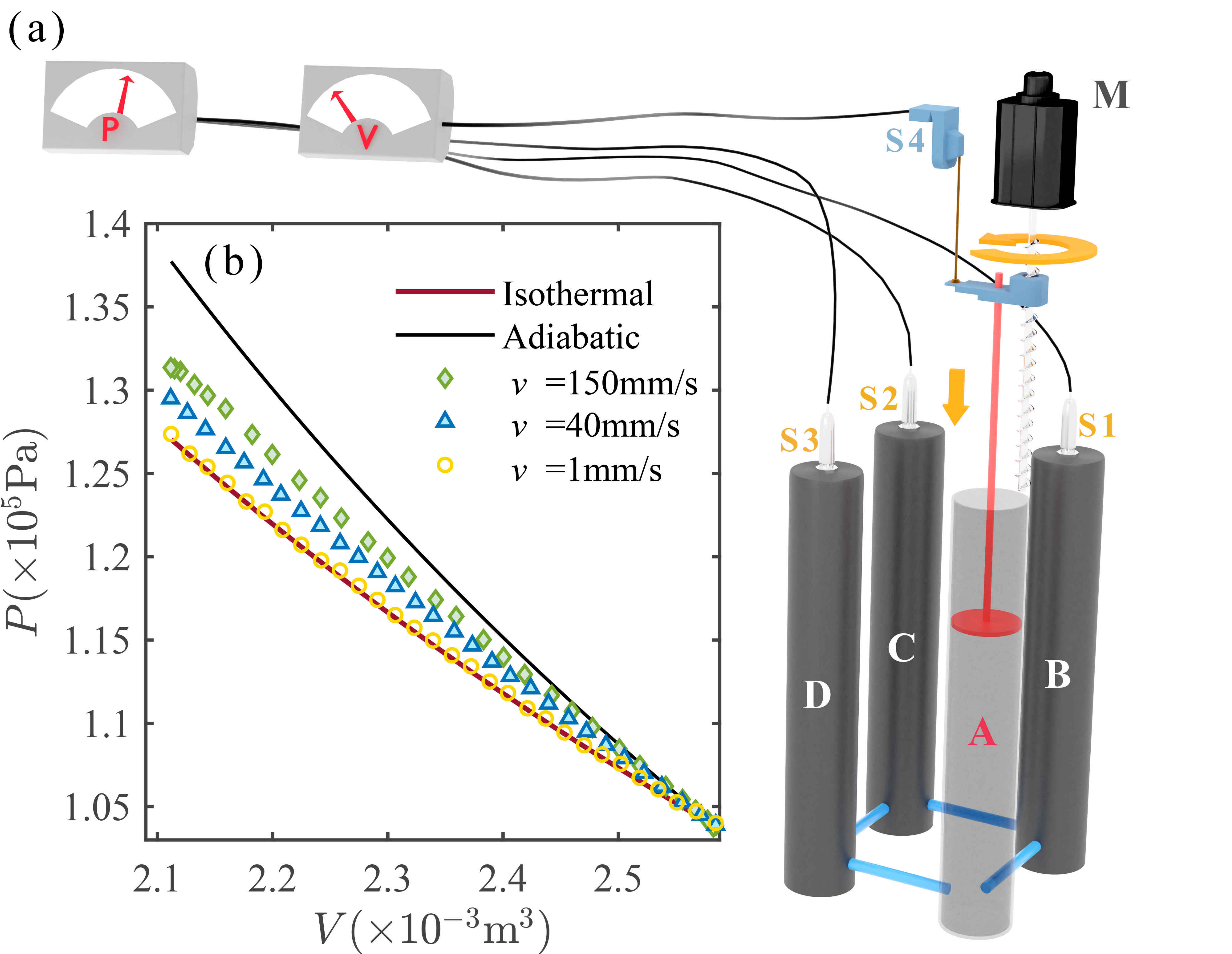}
\par\end{centering}
\centering{}\caption{\label{fig:Schematic} Experimental setup for measuring irreversible
entropy generation in the finite-time isothermal process. (a) Experimental
setup. The dry air is sealed in four connected cylinders A, B, C,
and D. The piston of A is propelled by the computer-controlled stepper
motor M to achieve the controlled compression of the gas. Three pressure
sensors S1, S2, and S3 are connected to the top of the three cylinders
B, C, and D respectively to measure the air pressure in the cylinders
$P(t)$. And the displacement of the piston $L\left(t\right)$ is
detected by a position sensor S4 to reveal the gas volume $V$. The
cylinders are immersed in the water bath with adjustable temperature.
(b) $P-V$ diagram of the gas under the bath temperature $T_{\mathrm{e}}=313.15$K
is illustrated in figure (b). The green diamonds, blue triangles,
and yellow circles are obtained for the piston speeds $150$mm/s,
$40$mm/s, and $1$mm/s, respectively. The red solid line shows the
theoretical quasi-static isothermal process, namely, $PV=\mathrm{const}$,
and the black solid line represents the adiabatic process with $PV^{\gamma}=\mathrm{const}$.
Here, $\gamma=1.4$ is the heat capacity ratio of the dry air \citep{Krause2004Determining}.}
\end{figure}

To our best knowledge, the direct verification of the $\mathscr{C}/\tau$
scaling was rarely explored, although the behavior of finite-time
heat engines has been studied in several experimental platforms \citep{MicrosizeHENatPhysics2011,ionHEPRL2012,Brantut2013,NanoscaleHEPRL2014,DynamicRelaxationNatTech2014,BrownianHENatPhys2015,Rossnagel2016,shortcutHEexperimentSciAdv2018}.
In this letter, we focus on experimental measuring IEG of dry air
via the work done in the finite-time isothermal process with designed
apparatus, and reveal the impact of control scheme on the IEG quantitatively.
In order to validate the $\mathscr{C}/\tau$ scaling, a controlled
apparatus in Fig. \ref{fig:Schematic}(a) is designed to measure the
state of the dry air, which is sealed in a compressible cylinder (A)
and three buffer cylinders (B, C, D), immersed in a temperature-controlled
water bath. A piston is installed in the cylinder A to compress the
air with a computer-controlled stepper motor M. By setting different
push programs, a controllable change in the volume of the gas over
time is achieved, i.e., $V\left(t\right)=V_{0}-\mathcal{A}L\left(t\right)$,
where $V_{0}=2.584\times10^{-3}\textrm{m}^{3}$ is the initial volume
of the gas, and $\mathcal{A}=1.963\times10^{-2}$$\textrm{m}^{2}$
is the cross sectional area of the cylinder A. The current setup allows
us to realize the finite-time isothermal process with different process
time $\tau$.

We firstly sketch the origin of the $\mathscr{C}/\tau$ scaling for
a general classic system, which contacts with a heat bath of constant
temperature $T_{\mathrm{e}}$. A control parameter, e.g., the volume
$V$ of the gas, is tuned from $t=0$ to the end of the process $t=\tau$.
In this process, with the endo-reversible assumption \citep{CA,endoreversibleHE,MinEntropyProd,Sahin1996},
IEG is written as \citep{MinEntropyProd}
\begin{equation}
\Delta S^{\left(\mathrm{ir}\right)}=\int_{0}^{\tau}\left(\frac{dQ_{\mathrm{s}}}{T_{\mathrm{s}}}+\frac{dQ_{\mathrm{e}}}{T_{\mathrm{e}}}\right),\label{eq:Sir}
\end{equation}
where $dQ_{\mathrm{s}}=-dQ_{\mathrm{e}}$ is the heat absorbed by
the system from the heat bath. The effective temperature $T_{\mathrm{s}}$
of the system generally varies with time in the control process. In
the condition of the quasi-static process with infinite control time
($\tau\rightarrow\infty$), the system is always in the thermal equilibrium
with $T_{\mathrm{s}}=T_{\mathrm{e}}$. For the long time $\tau$ in
comparison to the relaxation time $t_{\mathrm{r}}$ between the gas
and the heat bath, the system is in the linear irreversible region,
such that $T_{\mathrm{s}}$ is slightly deviated from the bath temperature,
namely $\left|T_{\mathrm{s}}-T_{\mathrm{e}}\right|/T_{\mathrm{e}}\ll1$.
The heat exchange rate between the system and bath follows the Newton's
law of cooling as

\begin{equation}
\frac{dQ_{\mathrm{s}}}{dt}=-\kappa(T_{\mathrm{s}}-T_{\mathrm{e}}),\label{eq:NLC}
\end{equation}
where $\kappa$ is the thermal conductance. Combining Eqs. (\ref{eq:Sir})
and (\ref{eq:NLC}), we obtain IEG as

\begin{equation}
\Delta S^{\left(\mathrm{ir}\right)}=\frac{\int_{0}^{1}J^{2}d\widetilde{t}}{\kappa T_{\mathrm{e}}^{2}\tau},\label{eq:Sir-general}
\end{equation}
where $J=dQ_{\mathrm{s}}/d\widetilde{t}$ is the heat flux, and $\widetilde{t}=t/\tau$
is the normalized time. The above equation shows the origin of $1/\tau$
scaling for the IEG.

For the current dry air system with volume compressed from $V_{0}$
to $V_{\mathrm{f}}$, the IEG is found proportional to the irreversible
work $W^{\left(\mathrm{ir}\right)}$ (IW) in the process under the
long time limit as follow
\begin{equation}
W^{\left(\mathrm{ir}\right)}=T_{\mathrm{e}}\Delta S^{(\mathrm{ir})}=\frac{P_{0}^{2}\left(V_{\mathrm{f}}-V_{0}\right)^{2}}{\kappa T_{\mathrm{e}}\tau},\label{eq:Sir-gas}
\end{equation}
where $P_{0}$ is the initial pressure of the dry air. The irreversible
work $W^{\left(\mathrm{ir}\right)}\left(\tau\right)=W\left(\tau\right)-W_{\mathrm{q}}$
is obtained by subtracting the work $W_{\mathrm{q}}=P_{0}V_{0}\ln\left(V_{0}/V_{\mathrm{f}}\right)$
in the quasi-static isothermal process from the work $W\left(\tau\right)=-\int_{0}^{\tau}PdV$
in the finite-time isothermal process. The detail of the current derivation
is shown in the Supplementary Materials. We will characterize the
irreversibility of the current system via the irreversible work, which
is a directly measurable quantity \citep{MinEntropyProd,andresen1984thermodynamics}
in our current setup.

\textit{Verification of $1/\tau$ scaling} - To measure the work $W(\tau)$,
we monitor the pressure $P=P(t)$ with three sensors, numbered S1,
S2 and S3 (range 0-0.15Mpa, accuracy 0.1\%) on the top of the three
cylinders B, C, and D respectively. The volume change $dV=\mathcal{A}dL\left(t\right)$
is measured through the piston position $L\left(t\right)$ with the
sensor S4 (range 0-0.3m, accuracy 0.1\%).

In the whole compression process, the four cylinders are immersed
in a large water bath ( volumn $100$L) with controllable temperature
(accuracy 0.5K). The internal equilibrium time of the gas is much
smaller than the relaxation time $t_{\mathrm{r}}$ that the gas is
always in the equilibrium state with temperature $T_{\mathrm{s}}$,
known as the endo-reversible \citep{CA,endoreversibleHE,MinEntropyProd,Sahin1996}.
In the current setup, $t_{\mathrm{r}}=1.942$s is measured in the
experiment with details explained in the Supplementary Materials.

\begin{figure}
\begin{centering}
\includegraphics[width=8cm]{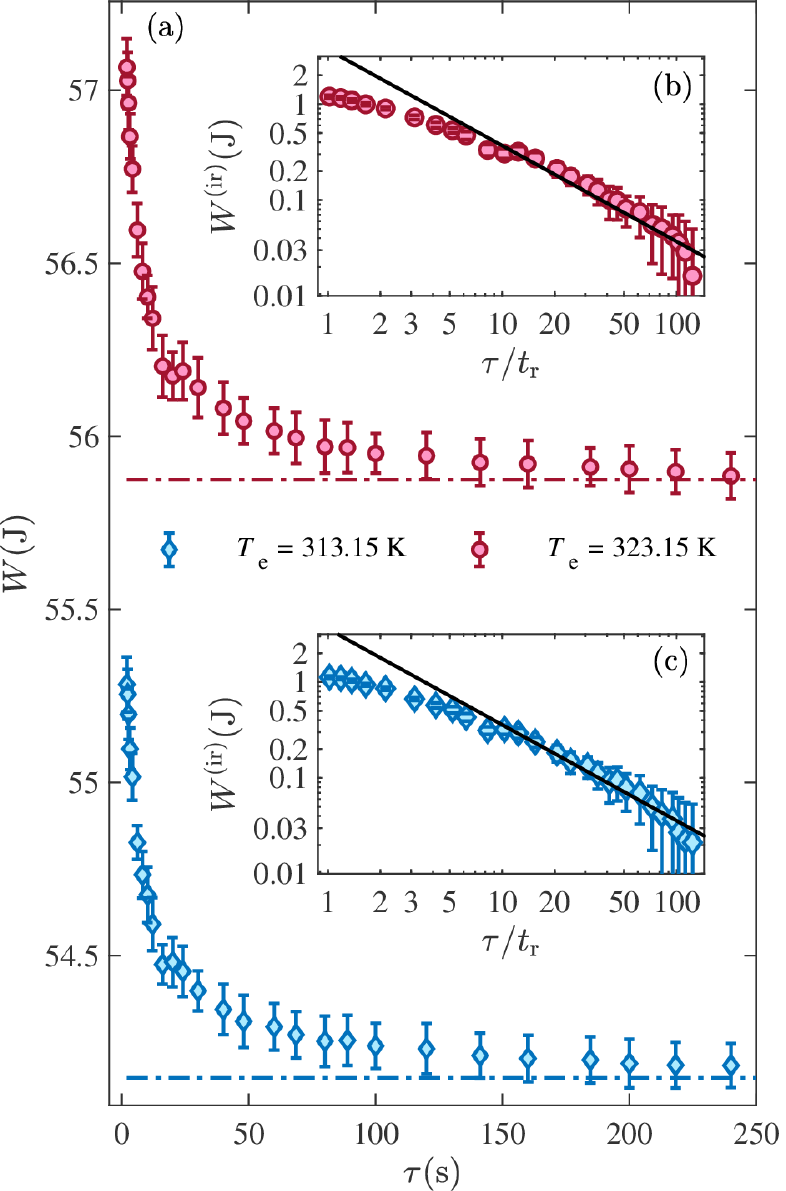}
\par\end{centering}
\centering{}\caption{\label{fig:Work-done-and}Work in the finite time isothermal process.
(a) Work done by the piston on the gas as the function of the process
time $\tau$. The experimental results are illustrated by the red
circled and blue diamond with the corresponding bath temperature $T_{\mathrm{e}}=323.15$K
and $T_{\mathrm{e}}=313.15$K respectively. The work $W_{\mathrm{q}}$
in the quasi-static process is marked by the red (blue) dash-dotted
line for $T_{\mathrm{e}}=323.15\textrm{K}\,(313.15\textrm{K})$. The
log-log plot of the irreversible work $W^{(\mathrm{ir})}$ as the
function of dimensionless time $\tau/t_{r}$ is illustrated in (b)
with $T_{\mathrm{e}}=323.15$K and (c) with $T_{\mathrm{e}}=313.15$K.
The corresponding theoretical result of Eq. (\ref{eq:Sir-gas}) is
represented by the black solid line.}
\end{figure}

The state of the dry air is illustrated via the P-V diagram in Fig.
\ref{fig:Schematic}(b). The pressure $P\left(t\right)$ is obtained
from the sensor S1 and the volume is measured by $V\left(t\right)=V_{0}-\mathcal{A}L\left(t\right)$
with $L(t)$ from the sensor S4. The total displacement of the piston
is $\Delta L=L\left(\tau\right)=240$mm. The sample frequency for
all sensors is set as $50\mathrm{Hz}$. In the plot, we show the P-V
diagram for the different piston speeds $v$, $150$mm/s (green diamond),
$40$mm/s (blue triangle), and $1$mm/s (yellow circle). It can be
seen from Fig. \ref{fig:Schematic}(b) that the slower the pushing
speed, the closer the curve is to the quasi-static isothermal process,
as shown by the red solid line. Conversely, the less heat exchange
between the gas and the heat bath for the fast push, and the $P-V$
curve is closer to the adiabatic process marked with the black solid
line ($PV^{\gamma}=\mathrm{const}$). Here $\gamma=1.4$ is the heat
capacity ratio for dry air \citep{Krause2004Determining}. The data
from pressure sensors S2 and S3 are illustrated in Supplementary Materials. 

By integrating the $P-V$ curve, we obtain the work done by the piston
to the gas as

\begin{equation}
W(\tau)=-\int_{0}^{\tau}P\left(t\right)\dot{V}\left(t\right)dt.
\end{equation}
The work as the function of the process time $\tau$ is illustrated
in Fig. \ref{fig:Work-done-and}(a), where the red circle and blue
diamond are obtained by setting the bath temperature $T_{\mathrm{e}}$=323.15K
and $T_{\mathrm{e}}$=313.15K respectively. Fig. \ref{fig:Work-done-and}(a)
shows that the work approaches a stable value, which matches the work
in the quasi-static isothermal process $W_{\mathrm{q}}$ (the dash-dotted
line). As shown with the log-log plot of the IW in Fig. \ref{fig:Work-done-and}(b)
and (c), in the long time region of $\tau\gg t_{r}$, the experimental
obtained IW is in good agreement with the theoretical prediction of
Eq. (\ref{eq:Sir-general}),which is represented by the black solid
line. Therefore, we validate the behavior that the IEG is inversely
proportional to the process time in the long-time region.

\begin{figure}
\includegraphics{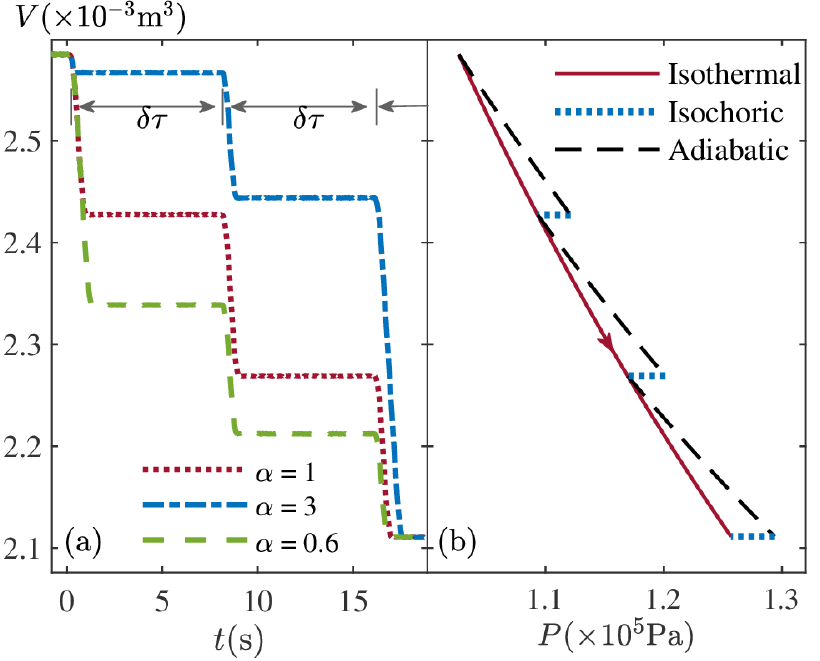}
\centering{}\caption{\label{fig:-diagram-of-DIP}The volume change and the $P-V$ diagram
in the discrete isothermal process with the 3-step case as an example.
(a) Volume changes with time in the 3-step DIP with different push
modes, where the step time is $\delta\tau=8$s. The piston is pushed
with $L_{i}=\left(i/3\right)^{\alpha}\Delta L,(i=1,2,3)$, where $L_{i}$
is the displacement of the piston after the end of the $i$-th step.
The gas volume $V_{i}=V_{0}-\mathcal{A}L_{i}$ being tuned sub-linearly
($\alpha=0.6$), linearly ($\alpha=1$), and super-linearly ($\alpha=3$)
are illustrated by the green dashed line, red dotted line, and blue
dash-dotted line respectively. (b) P-V diagram of the 3-step DIP.
Series of adiabatic (black dashed line) and isochoric (blue dotted
line) processes are used to approach a finite-time isothermal process
(red solid line). In the $i$-th ($i=1,2,3$) step, the gas volume
is firstly compressed from $V_{i}$ to $V_{i+1}$ adiabatically, then
the gas isochornically relaxes to the thermal equilibrium state with
the same temperature $T_{e}$ as the water bath. The experimental
$P-V$ diagram is shown in the Supplementary Materials.}
\end{figure}

\textit{Effect of the control scheme} - With the above compression
process at the constant speed, we have validated the $1/\tau$ scaling
of the IEG via the measurement of the IW. As predicted in the previous
study \citep{MinEntropyProd,yhmaoptimalcontrol}, the coefficient
in the $1/\tau$ scaling relation of IW not only is determined by
the system parameters of the working substance and heat bath, but
also relates to the specific way how the state of the working substance
is tuned. In the following experiment, we will show the impact of
different control schemes on IW with our setup via a discrete isothermal
process \citep{andresen1984thermodynamics}.

The discrete isothermal process, introduced by Andresen et al. in
Ref. \citep{Andresen1977}, is an effective approach to optimize the
finite-time Carnot engine. Since then, the discrete step thermodynamic
process have also been used to study of different thermodynamic issues,
such as work distribution \citep{Quan2008}, thermodynamic length
\citep{Crooks2007}, and optimization of quantum heat engines \citep{Geva1992,yhmaoptimalcontrol}.
The basic idea of the discrete step isothermal process (DIP) is to
use a series of adiabatic and isochoric processes to construct a finite-time
isothermal process. The discrete isothermal process has two obvious
advantages, theoretically the state of the working substance can be
analytically solved and experimentally the work and the heat exchange
are separated for direct measurement.

In our setup, the piston is rapidly pushed in the $i$-th step to
the position $L_{i}$ ($i=1,2,...,M$) to form an adiabatic process,
and then relaxes to thermal equilibrium through the isochoric process
with duration $\delta\tau$, as shown in Fig. \ref{fig:-diagram-of-DIP}(a).
The initial (final) piston position is $L_{0}=0$ ($L_{M}=\Delta L$).
For clarity, we show three control schemes with different $\alpha$
with total step number $M=3$ and duration time $\delta\tau=8$s as
an example. At the beginning of the each adiabatic process, the gas
maintains the same temperature as the water bath, since $\delta\tau$
is larger than the relaxation time $t_{\mathrm{r}}$ that $\exp[-\delta\tau/t_{\mathrm{r}}]\ll1$.
We define the average speed of the piston in one step as $v_{i}=(L_{i}-L_{i-1})/\delta\tau$.
The stepper motor can be set to push the piston $L_{i}$ with a power
function $L_{i}=\left(i/M\right)^{\alpha}\Delta L$ in the $i$-th
step. 

For the discrete isothemal process involving $M\gg1$ steps, the IW
of the system is explicitly written as {[}See Supplementary Materials
for detailed derivation{]} 

\begin{equation}
W^{\left(\mathrm{ir}\right)}=\frac{\Lambda\Theta}{M},\label{eq:Sir-dis}
\end{equation}
where $\Theta=\left(\gamma-1\right)P_{0}\left(V_{\mathrm{f}}-V_{0}\right)^{2}/(2V_{0})$
relates to the initial and final state of the system. And $\Lambda=\left\langle v^{2}\right\rangle /\left\langle v\right\rangle ^{2}$,
characterizing the speed fluctuation of the piston, is determined
by the control scheme of the stepper motor with $\left\langle v^{2}\right\rangle \equiv\sum_{i=1}^{M}v_{i}^{2}/M$
and $\left\langle v\right\rangle =\sum_{i=1}^{M}v_{i}/M$. The current
general formula in Eq. ( \ref{eq:Sir-dis}) recovers the result for
piston compressed with the constant speed noticing $\Lambda=1$. With
the fixed process time $\tau=M\delta\tau$, any control scheme under
power function \citep{yhmaoptimalcontrol} results in the larger $\Lambda>1$,
which in turn induces the larger IW $W^{\left(\mathrm{ir}\right)}$
than that with the constant speed.

With the current setup, we can experimentally demonstrate the effect
of the control function on the IW. The control functions are realized
by the different power indexes $\alpha$. The volume change of the
gas in a 3-step DIP is illustrated in Fig. \ref{fig:-diagram-of-DIP}(a),
where the green dashed line, red dotted line, and blue dash-dotted
line relate to the piston been pushed sub-linearly, linearly, and
super-linearly, respectively. The schematic $P-V$ curve for the DIP
is illustrated in Fig. \ref{fig:-diagram-of-DIP}(b). 

\begin{figure}
\begin{raggedright}
\includegraphics{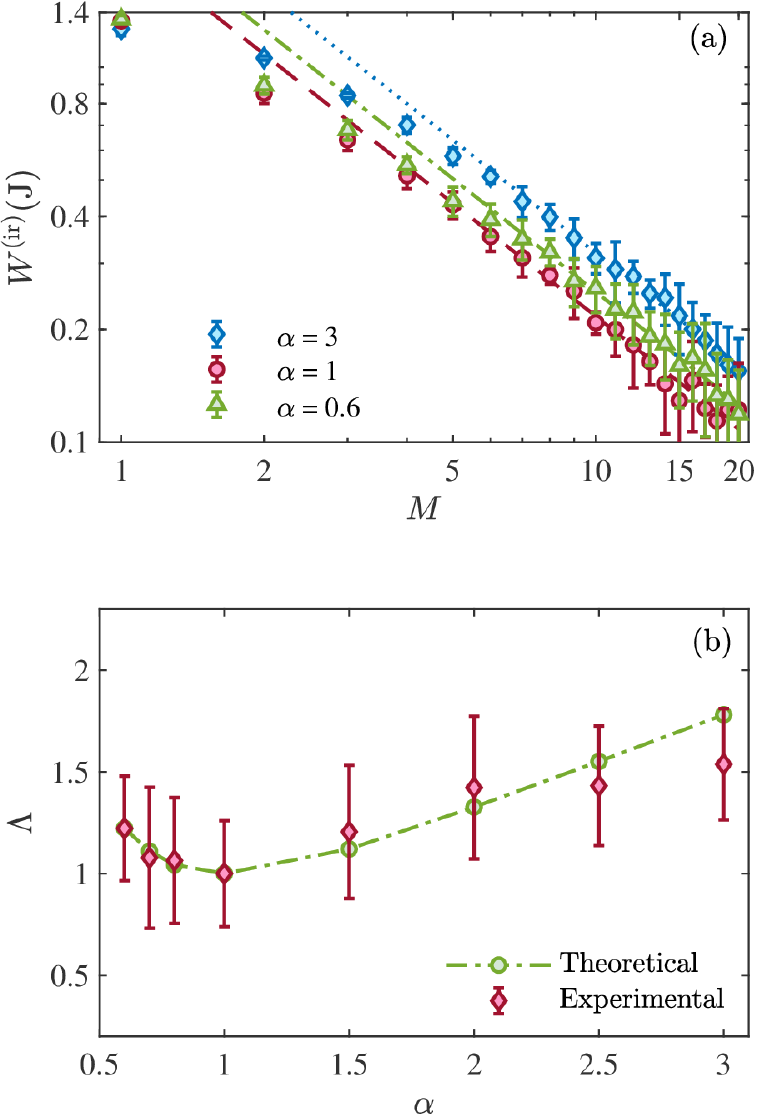}
\par\end{raggedright}
\caption{\label{fig:Irreversible-entropy-generation}Irreversible work with
different piston push schemes of the discrete isothermal process.
The temperature of the water bath is $T_{\mathrm{e}}=313.15$K . (a)
log-log plot of irreversible work as the function of step number $M$.
We demonstrate the $1/M$ scaling for three control functions with
$\alpha=3$ (the blue diamond), $0.6$ (green triangle), and $1$
(red circle). (b) The obtained parameter $\Lambda$ in Eq. (\ref{eq:Sir-dis})
as the function of power index $\alpha$. The experiment results,
represented by the red diamond, are obtained by fitting the relation
of $W^{\left(\mathrm{ir}\right)}\sim1/M$. The theoretical curve in
Eq. (\ref{eq:Sir-dis}) is plotted with the green dash-dotted line
as a comparison.}
\end{figure}

The irreversible work done in DIP is obtained by integrating area
under the P-V curve, and illustrated in Fig. \ref{fig:Irreversible-entropy-generation}(a)
as a function of the total step number $M$ for three different power
indexes $\alpha=0.6$ (green triangle), $1.0$ (red circle) and $3.0$
(blue diamond). Each data points have been averaged with 20 repeats.
The corresponding dashed lines show the fitting with the theoretical
result in Eq. (\ref{eq:Sir-dis}). At the large-$M$ region, the IW
is inversely proportional to $M$, namely, inversely proportional
to the total time $\tau$.

To show the dependence of the IW on the control function, we plot
the coefficient $\Lambda$ of the $1/M$ scaling in Eq. (\ref{eq:Sir-dis})
as a function of the index $\alpha$ in Fig. \ref{fig:Irreversible-entropy-generation}(b).
The experimental data for coefficient $\Lambda$, shown as diamonds
in Fig. \ref{fig:Irreversible-entropy-generation}(b), is obtained
by fitting curves in \ref{fig:Irreversible-entropy-generation}(a)
with Eq. (\ref{eq:Sir-dis}) for different $\alpha$ at large step
number $M$. The theoretical result of Eq. (\ref{eq:Sir-dis}) is
shown as the green circle in Fig. \ref{fig:Irreversible-entropy-generation}.
The figure shows the agreement between the theoretical result and
the experimental data. The experimental data shows a minimum irreversible
work at $\alpha=1$. We conclude that within the set of power function,
the minimal IW is achieved with the linearly control function \citep{yhmaoptimalcontrol},
namely $\alpha=1$ as shown in Fig. \ref{fig:Irreversible-entropy-generation}.

With the dependence of the control function $\Lambda$, we can control
the IW of the system by different schemes of compression to adjust
the power and efficiency of the heat engine \citep{yhmaoptimalcontrol}.
Experimentally, such tunning of irreversible entropy generation via
adjusting the mode of operation is meaningful for the design of heat
engine with high output power and efficiency.

\textit{Conclusion-}We have designed the apparatus with the cylinder-gas
system to validate the theoretically predicted $1/\tau$ scaling of
irreversible entropy generation in the finite-time thermodynamics.
Our experiment for the first time directly shows that the irreversible
entropy generation, obtained by measuring the irreversible work, is
inversely proportional to the process time $\tau$ in the long-time
region {[}Fig. \ref{fig:Work-done-and}(b){]}, namely, $\Delta S^{\left(\mathrm{ir}\right)}\propto1/\tau$.
More importantly, we demonstrated the proportional relationship between
IEG and the speed fluctuation of the piston in different gas compression
schemes for the discrete isothermal process. Specifically, we verified
the minimal IEG can be achieved by pushing the piston linearly within
the set of the power control functions. This provides a feasible and
convenient solution for the optimization of the actual heat engine
by applying different control schemes to the work substance in different
processes of the thermodynamic cycle. 

The similar detection of the irreversible work can also be realized
in quantum system, such as trapped ions \citep{ionHEPRL2012,an2015experimental,Rossnagel2016},
NMR system \citep{de2019efficiency} and superconducting circuit systems
\citep{Giazotto2006,uzdin2019experimental}. The generalization of
the current measurement in quantum regime could potentially shows
the influence of coherence on these thermodynamic quantities \citep{Brandner2017,Su2018,Camati2019}.

Yu-Han Ma is grateful to Hong Yuan and Jin-Fu Chen for helpful discussions.
This work is supported by the NSFC (Grants No. 11534002 and No. 11875049),
the NSAF (Grant No. U1730449 and No. U1530401), and the National Basic
Research Program of China (Grants No. 2016YFA0301201 and No. 2014CB921403).
H.D. also thanks The Recruitment Program of Global Youth Experts of
China.

\bibliographystyle{apsrev}
\bibliography{ExpTestScalingV7_3_20191029.bbl}
\end{document}


\title{Supplementary Materials to ``Experimental validation of the $1/\tau$-scaling
entropy generation in finite-time thermodynamics with dry air''}
\author{Yu-Han Ma}
\address{Beijing Computational Science Research Center, Beijing 100193, China}
\address{Graduate School of China Academy of Engineering Physics, No. 10 Xibeiwang
East Road, Haidian District, Beijing, 100193, China}
\author{Ruo-Xun Zhai}
\address{Beijing Normal University, Beijing 100875, China}
\address{Graduate School of China Academy of Engineering Physics, No. 10 Xibeiwang
East Road, Haidian District, Beijing, 100193, China}
\author{Chang-Pu Sun}
\address{Beijing Computational Science Research Center, Beijing 100193, China}
\address{Graduate School of China Academy of Engineering Physics, No. 10 Xibeiwang
East Road, Haidian District, Beijing, 100193, China}
\author{Hui Dong}
\email{hdong@gscaep.ac.cn}
\address{Graduate School of China Academy of Engineering Physics, No. 10 Xibeiwang
East Road, Haidian District, Beijing, 100193, China}

\maketitle

\onecolumngrid
\setcounter{equation}{0}
\setcounter{figure}{0}
\setcounter{table}{0}
\setcounter{page}{1}
\renewcommand{\theequation}{S\arabic{equation}}
\renewcommand{\thefigure}{S\arabic{figure}}
\renewcommand{\bibnumfmt}[1]{[S#1]}
\renewcommand{\citenumfont}[1]{S#1}

This document is devoted to providing the detailed derivations and
the supporting discussions to the main content in the Letter.

\section{Irreversible entropy generation and irreversible work of dry air}

In this section we derive the relation between the irreversible work
and the irreversible entropy generation for the current system with
dry air. The irreversible entropy generation is generally defined
as

\begin{align}
\Delta S^{\mathrm{(ir)}} & =\int_{0}^{\tau}\left(\frac{dQ}{T_{\mathrm{s}}}-\frac{dQ}{T_{\mathrm{e}}}\right)\\
 & =\int_{0}^{\tau}\frac{dU+PdV}{T_{\mathrm{s}}}-\frac{1}{T_{\mathrm{e}}}\int_{0}^{\tau}\left(dU+PdV\right)\\
 & =\int_{0}^{\tau}\frac{C_{V}dT+\frac{nRT_{\mathrm{s}}}{V}dV}{T_{\mathrm{s}}}-\frac{1}{T_{\mathrm{e}}}\int_{0}^{\tau}\left(C_{V}dT+PdV\right)\\
 & =C_{V}\ln\left(\frac{T_{\mathrm{s}}\left(\tau\right)}{T_{e}}\right)-C_{V}\frac{T_{\mathrm{s}}\left(\tau\right)-T_{e}}{T_{e}}-\frac{\int_{0}^{\tau}PdV}{T_{e}}+nR\ln\left(\frac{V_{\mathrm{f}}}{V_{0}}\right).\label{eq:Sir-W}
\end{align}
Under the long time limit with $\left|T_{\mathrm{s}}-T_{\mathrm{e}}\right|/T_{\mathrm{e}}\ll1$,
the irreversible entropy generation $\Delta S^{\mathrm{(ir)}}$ is
simplified to the first order of $(T_{\mathrm{s}}-T_{\mathrm{e}})/T_{\mathrm{e}}$
as 
\begin{equation}
\Delta S^{\mathrm{(ir)}}=\frac{W-W_{\mathrm{q}}}{T_{\mathrm{e}}},
\end{equation}
where $W(\tau)=-\int_{0}^{\tau}PdV$ is the total work done by the
piston to the gas during the finite-time process and $W_{\mathrm{q}}=nRT_{\mathrm{e}}\ln(V_{0}/V_{\mathrm{f}})$
is the work done for quasi-static process. Here, $n$ is the number
of moles and $R$ is the ideal gas constant. The irreversible work
is defined as $W^{\mathrm{(ir)}}=W-W_{\mathrm{q}}$, which is connected
to the entropy generation via 
\begin{equation}
W^{\mathrm{(ir)}}=T_{\mathrm{e}}\Delta S^{\mathrm{(ir)}}.
\end{equation}
 This simplification allows the direct measurement of the irreversibility
via irreversible work $W^{\mathrm{(ir)}}$ in the current setup.

\section{Measure the relaxation time $t_{r}$}

\begin{figure}[!htbp]
\centering
\includegraphics{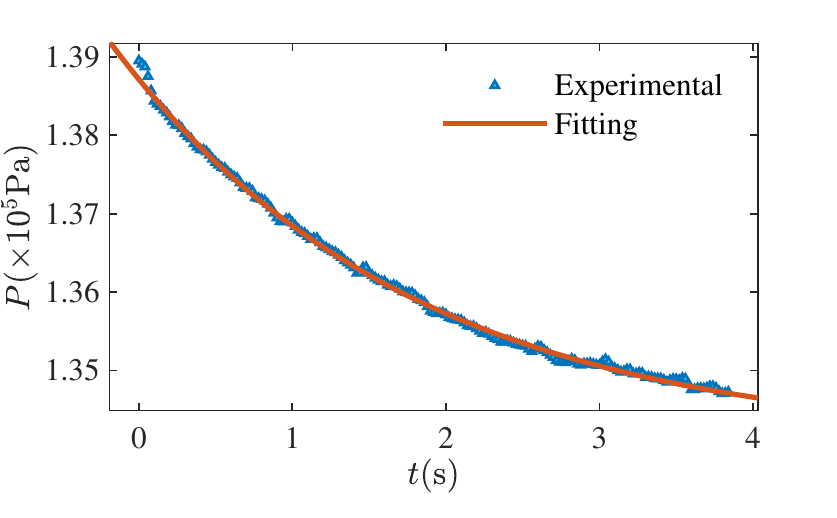}
\caption{\label{fig:relaxationtime}The pressure change after the fast compression.
The blue triangles show the experimental data from pressure sensor
S1, and the red line illustrates the fitting with the curve in Eq.
(\ref{eq:Pt}). The relaxation time $t_{r}=1.942$s is obtained by
fitting the experimental data.}
\end{figure}

In the data analysis, the relaxation time $t_{r}$ can be directly
determined via the relaxation process. During the isochornic process
with the pressure relaxation, the change of the internal energy of
the gas is caused by the heat exchange

\begin{equation}
\frac{dU}{dt}=\frac{dQ_{\mathrm{s}}}{dt}=-\kappa(T_{\mathrm{s}}-T_{\mathrm{e}}).
\end{equation}
Combining with internal energy equation $dU=C_{V}dT$, we have the
explicit evolution of the temperature as

\begin{equation}
T_{\mathrm{s}}\left(t\right)=T_{\mathrm{e}}+\left[T_{\mathrm{s}}\left(0\right)-T_{\mathrm{e}}\right]e^{-t/t_{r}},
\end{equation}
where, $t_{\mathrm{r}}=C_{V}/\kappa$ is the relaxation time. The
dynamical change of the temperature is directly reflected through
the change of the pressure via the ideal gas equation as $P\left(t\right)=nRT_{\mathrm{s}}(t)$.
To measure the relaxation time, we compress the sealed gas with the
maximum speed to the final volume $V_{\mathrm{f}}$, and measure the
pressure change $P(t)$. The measured data, shown in Fig. \ref{fig:relaxationtime},
is fitted with the curve

\begin{align}
P\left(t\right) & =\frac{nR}{V}[T_{\mathrm{e}}+\left[T_{\mathrm{s}}\left(0\right)-T_{\mathrm{e}}\right]e^{-t/t_{\mathrm{r}}}].\label{eq:Pt}
\end{align}
The experimental fitting results in the relaxation time $t_{\mathrm{r}}=1.942\mathrm{s}$.

\section{Irreversible entropy generation in discrete isothermal process}

\begin{figure}
\begin{centering}
\includegraphics{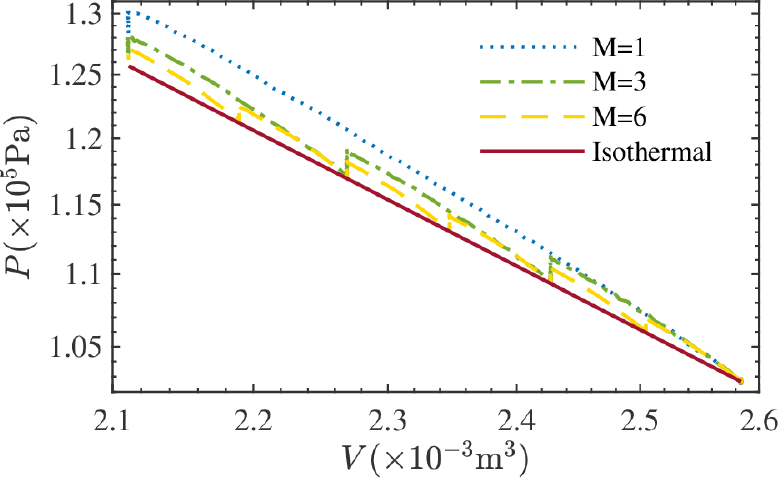}
\par\end{centering}
\caption{\label{fig:-diagram-of-PV-DIP}$P-V$ diagram of the gas in discrete
isothermal process. $P-V$ diagram with different step number $M$.
The blue dotted line, green dash-dotted line, and yellow dashed line
relates to $M=1$, $M=3$, and $M=6$, respectively. The isothermal
line of the gas is illustrated by the red solid line.}
\end{figure}

In this section, we provide the detailed derivation of Equation (6)
in the letter. The discrete isothermal process (DIP) consisting of
a series of adiabatic processes and isochoric processes is used to
approach the quasi-static isothermal process. The experimental obtained
$P-V$ diagram of the gas in DIP with different step number $M$ are
illustrated in Fig. \ref{fig:-diagram-of-PV-DIP}.

The work done by the piston to the gas in the $i$-th step comes only
from the $i$-th adiabatic process as

\begin{equation}
W_{i}=-\int_{V_{i-1}}^{V_{i}}P_{i}\left(V\right)dV,
\end{equation}
where the pressure follows the adiabatic equation of idea gas as

\begin{equation}
P_{i}(V)=\frac{P_{i}^{0}V_{i-1}^{\gamma}}{V^{\gamma}},V\in\left[V_{i-1},V_{i}\right],
\end{equation}
where $\gamma=C_{P}/C_{V}$ is the heat capacity ratio of the gas.
Initially at $i$-th adiabatic step, the gas is in equilibrium with
the bath i.g., $T_{i}=T_{\mathrm{e}}$. And the pressure is $P_{i}^{0}=nk_{B}T_{i}/V_{i-1}$.
The work done at the $i$-th adiabatic process is obtained explicitly
by integration,

\begin{align}
W_{i} & =-\int_{V_{i-1}}^{V_{i}}\frac{P_{i}^{0}V_{i-1}^{\gamma}}{V^{\gamma}}dV=\frac{nRT_{\mathrm{e}}}{1-\gamma}\left[\left(\frac{V_{i}}{V_{i-1}}\right)^{1-\gamma}-1\right].\label{eq:Wi}
\end{align}
And the total work is the summation over all the adiabatic process

\begin{equation}
\Delta W=\sum_{i=1}^{M}W_{i}=-\frac{nRT_{\mathrm{e}}}{1-\gamma}\sum_{i=1}^{M}\left[\left(\frac{V_{i}}{V_{i-1}}\right)^{1-\gamma}-1\right].\label{eq:exact}
\end{equation}
A similar result was reported in Ref. \citep{Andresen1977}, where
the authors used isobaric processes instead of adiabatic processes.

The volume of the sealed gas is controlled by the piston via the function

\begin{equation}
V_{i}=V_{0}-\mathcal{A}L_{i},
\end{equation}
where $\mathscr{\mathcal{A}}$ is the cross section of the piston.
With the control, the work for each adiabatic process of Eq. (\ref{eq:Wi})
becomes 

\begin{align}
W_{i} & =-\frac{nRT_{\mathrm{e}}}{1-\gamma}\left[\left(\frac{V_{0}-\mathcal{A}L_{i}}{V_{0}-\mathcal{A}L_{i-1}}\right)^{1-\gamma}-1\right]\\
 & =-\frac{nRT_{\mathrm{e}}}{1-\gamma}\left[\left(1+\frac{L_{i-1}-L_{i}}{V_{0}/\mathcal{A}-L_{i-1}}\right)^{1-\gamma}-1\right]\\
 & \approx-\frac{nRT_{\mathrm{e}}}{1-\gamma}\left[\left(1-\gamma\right)\frac{L_{i-1}-L_{i}}{V_{0}/\mathcal{A}-L_{i-1}}-\frac{\gamma\left(1-\gamma\right)}{2}\left(\frac{L_{i-1}-L_{i}}{V_{0}/\mathcal{A}-L_{i-1}}\right)^{2}\right]\\
 & =\frac{nRT_{\mathrm{e}}v_{i}\delta\tau}{V_{0}/\mathcal{A}-L_{i-1}}+\frac{\gamma nRT_{\mathrm{e}}}{2}\left(\frac{v_{i}\delta\tau}{V_{0}/\mathcal{A}-L_{i-1}}\right)^{2},
\end{align}
where $v_{i}\equiv(L_{i}-L_{i-1})/\delta\tau$ is the average speed
of the piston in the $i$-th step. Then, by keeping up to the second
order of $\mathcal{A}L_{i}/V_{0}$, we obtain the work done in the
whole process as

\begin{align}
W & =\sum_{i=1}^{M}\left[\frac{nRT_{\mathrm{e}}v_{i}\delta\tau}{V_{0}/\mathcal{A}-L_{i}}+\frac{\gamma nRT_{\mathrm{e}}}{2}\left(\frac{v_{i}\delta\tau}{V_{0}/\mathcal{A}-L_{i}}\right)^{2}\right]\\
 & \approx\frac{nRT_{\mathrm{e}}\mathcal{A}}{V_{0}}\sum_{i=1}^{M}v_{i}\delta\tau\left(1+\frac{\mathcal{A}L_{i}}{V_{0}}\right)+\frac{\gamma nRT_{\mathrm{e}}}{2V_{0}^{2}}\sum_{i=1}^{M}\left(v_{i}\delta\tau\right)^{2}\label{eq:arbitrary}\\
 & =\frac{nRT_{\mathrm{e}}\mathcal{A}}{V_{0}}\sum_{i=1}^{M}v_{i}\delta\tau+\frac{nRT_{\mathrm{e}}\mathcal{A}^{2}}{V_{0}^{2}}\sum_{i=1}^{M}v_{i}\delta\tau\sum_{j=1}^{i}v_{j}\delta\tau+\frac{\gamma nRT_{\mathrm{e}}\mathcal{A}^{2}}{2V_{0}^{2}}\sum_{i=1}^{M}\left(v_{i}\delta\tau\right)^{2}\\
 & =\frac{nRT_{\mathrm{e}}\mathcal{A}}{V_{0}}\sum_{i=1}^{M}v_{i}\delta\tau+\frac{nRT_{\mathrm{e}}\mathcal{A}^{2}}{V_{0}^{2}}\left[\left(\sum_{i=1}^{M}v_{i}\delta\tau\right)^{2}-\sum_{i=1}^{M}\left(v_{i}\delta\tau\right)^{2}\right]+\frac{\gamma nRT_{\mathrm{e}}\mathcal{A}^{2}}{2V_{0}^{2}}\sum_{i=1}^{M}\left(v_{i}\delta\tau\right)^{2}\\
 & =\frac{nRT_{\mathrm{e}}\mathcal{A}L_{M}}{V_{0}}+\frac{nRT_{\mathrm{e}}\mathcal{A}^{2}L_{M}^{2}}{V_{0}^{2}}+\frac{\left(\gamma-1\right)nRT_{\mathrm{e}}\mathcal{A}^{2}}{2V_{0}^{2}}\sum_{i=1}^{M}\left(v_{i}\delta\tau\right)^{2}\\
 & =\frac{nRT_{\mathrm{e}}\left(V_{\mathrm{f}}-V_{0}\right)}{V_{0}}+\frac{nRT_{\mathrm{e}}\left(V_{\mathrm{f}}-V_{0}\right)^{2}}{V_{0}^{2}}+\frac{\left(\gamma-1\right)nRT_{\mathrm{e}}\left(V_{\mathrm{f}}-V_{0}\right)^{2}}{2V_{0}^{2}}\frac{\sum_{i=1}^{M}\left(v_{i}\right)^{2}}{\left(\sum_{i=1}^{M}v_{i}\right)^{2}}.\label{eq:W-DIP}
\end{align}
Here, $L_{M}=\delta\tau\sum_{i=1}^{M}v_{i}$ and $V_{\mathrm{f}}-V_{0}=\mathcal{A}L_{M}$
are respectively the total displacement of the piston and the change
of the gas volume in the whole process. Note that the first two terms
of Eq. (\ref{eq:W-DIP}) is just the Taylor's expansion of the work
done in quasi-static isothemal process $W_{\mathrm{q}}=-nRT_{\mathrm{e}}\ln(V_{\mathrm{f}}/V_{0})$
up to $(\Delta V/V_{0})^{2}$. Consequently, the irreversible work
$W^{\mathrm{(ir)}}=W-W_{\mathrm{q}}$ is given by

\begin{align}
W^{\mathrm{(ir)}} & =\frac{\left(\gamma-1\right)nRT_{\mathrm{e}}\left(V_{\mathrm{f}}-V_{0}\right)^{2}}{2V_{0}^{2}}\frac{\sum_{i=1}^{M}\left(v_{i}\right)^{2}}{\left(\sum_{i=1}^{M}v_{i}\right)^{2}}\\
 & =\frac{\left(\gamma-1\right)P_{0}\left(V_{\mathrm{f}}-V_{0}\right)^{2}}{2MV_{0}}\frac{\left\langle v^{2}\right\rangle }{\left\langle v\right\rangle ^{2}}
\end{align}
where $\left\langle v^{2}\right\rangle =\left[\sum_{i=1}^{M}v_{i}^{2}\right]/M$,
and $\left\langle v\right\rangle =\left(\sum_{i=1}^{M}v_{i}\right)/M$.
With the definitions $\Lambda\equiv\left\langle v^{2}\right\rangle /\left\langle v\right\rangle ^{2}$,
and $\Theta\equiv P_{0}\left(V_{\mathrm{f}}-V_{0}\right)^{2}/\left(2V_{0}\right)$,
the irreversible work in discrete isothermal process of Eq. (6) in
the main context is obtained.

\section{Additional results from pressure sensors S2 and S3}

\begin{figure}[!htbp]
\includegraphics{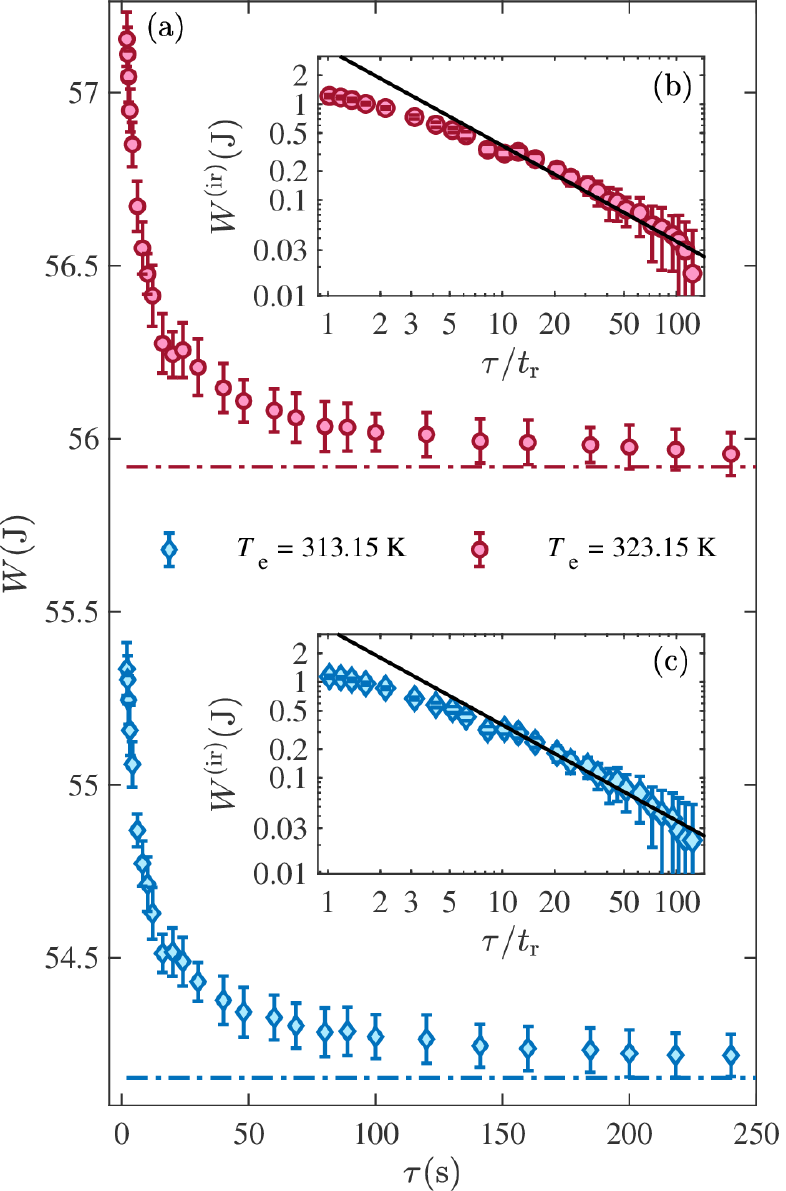}\includegraphics{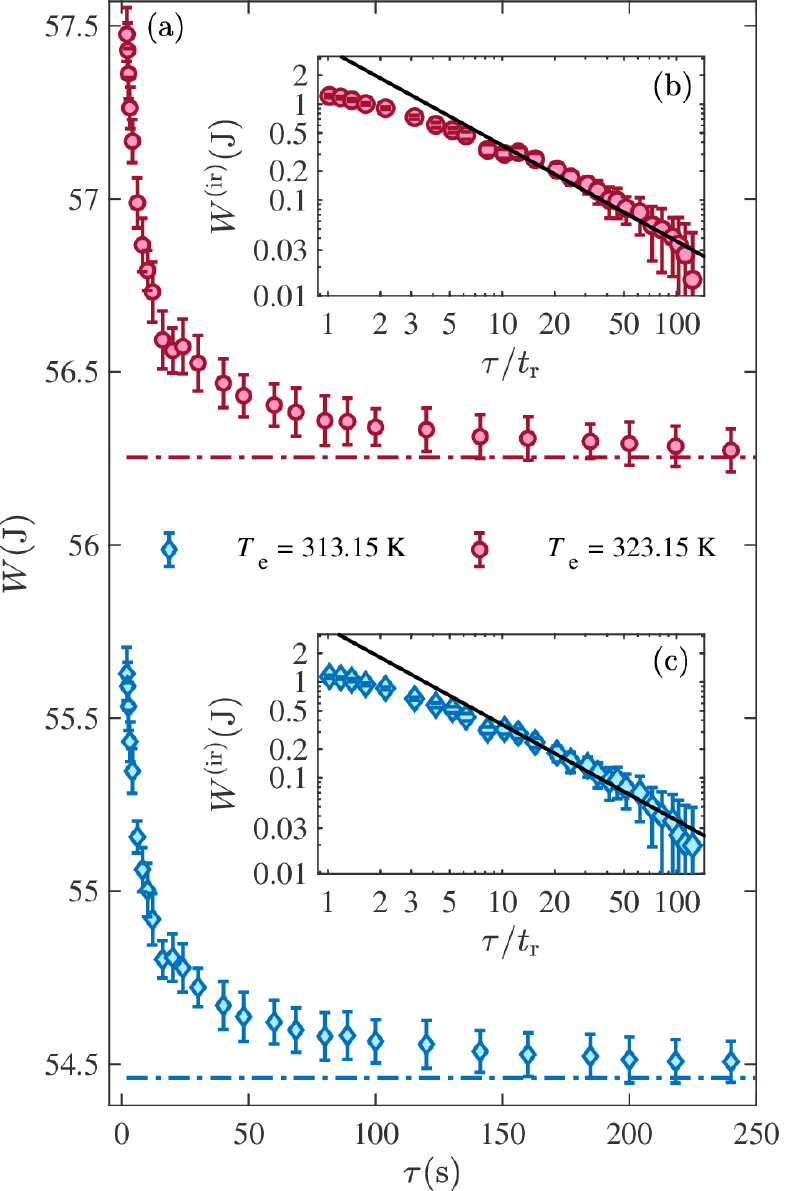}
\caption{\label{fig:linearS2S3}The irreversible work measured in the linear
control scheme with sensors S2 (left panel) and S3 (right panel).}
\end{figure}

\begin{figure}[!htbp]
\includegraphics{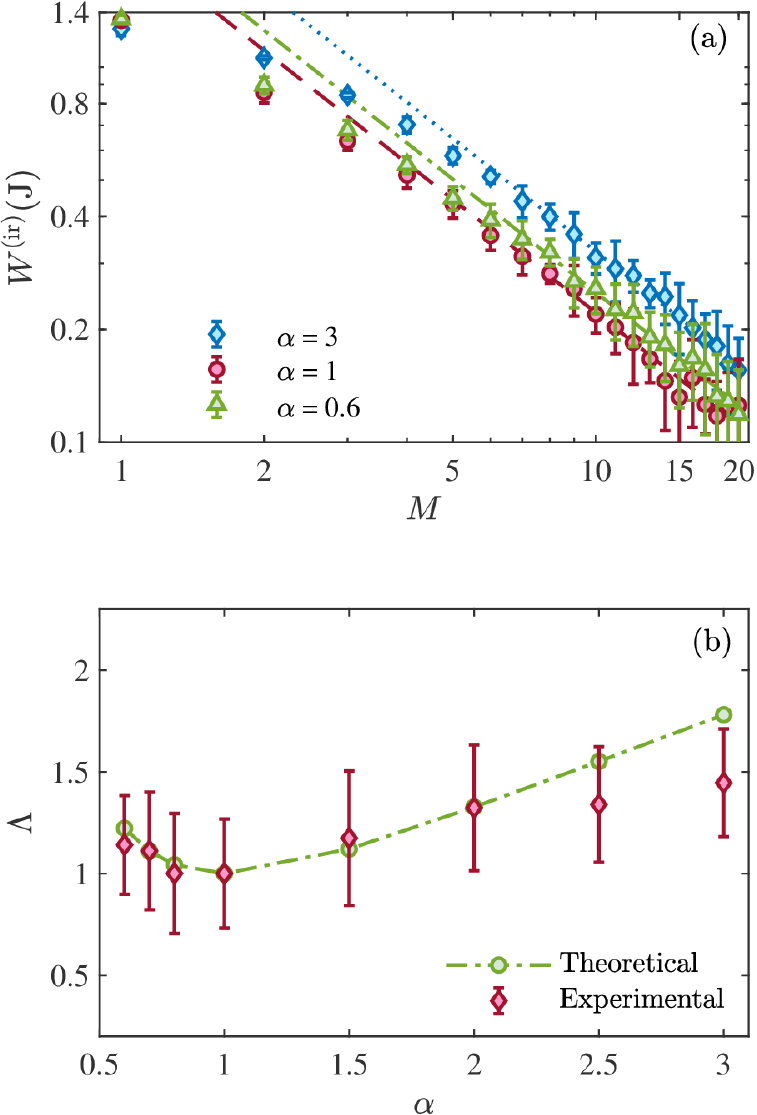}\includegraphics{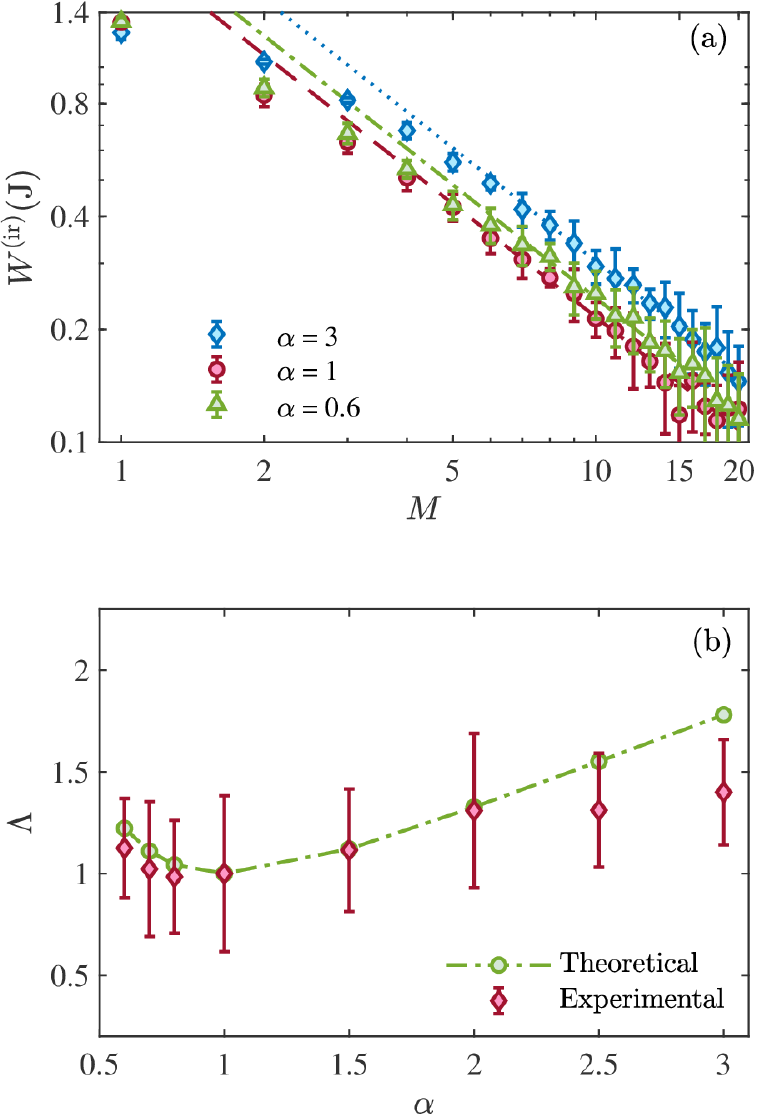}
\caption{\label{fig:dipS2S3T40}The irreversible work for the discrete process
and the coefficient $\Lambda$ estimated from experimental data with
sensors S2 (left panel) and S3 (right panel) at temperature $T_{\mathrm{e}}=313.15\mathrm{K}$.}
\end{figure}

In addition to the experimental data from pressure sensor S1 in the
main context, we show the results from the pressure sensor S2 and
S3 as following in Figures \ref{fig:linearS2S3}, and \ref{fig:dipS2S3T40}.
These figures illustrate mainly the data at the temperature at $T_{\mathrm{e}}=313.15\mathrm{K}$.
Similar figures for $T_{\mathrm{e}}=323.15\mathrm{K}$ can be obtained
upon request.

\bibliographystyle{apsrev4-1}
\bibliography{SupplementaryMaterialsV4_3_20191029.bbl}